\documentclass[11pt]{article}
\usepackage{amsmath, amssymb, amstext}
\def\mxth{\mathsurround=0pt }
\def\xversim#1#2{\lower2.pt\vbox{\baselineskip0pt \lineskip-.5pt
\ialign{$\mxth#1\hfil##\hfil$\crcr#2\crcr\sim\crcr}}}

\begin{document}
\begin{titlepage}
\begin{flushright}
{~}\\[12mm]
ITP-UU-05/33 \\
SPIN-05/27\\[3mm]
August 2005
\end{flushright}
\vskip 20mm
\begin{center}
  {\Large {\bf Supersymmetric Black Holes }}\footnote[1]{ To be
    published in the proceedings of the Symposium in honor of the 70th
    birthday of Julius Wess, M\"unchen, 10-11 January, 2005.}
\end{center}
\vskip 6mm

\begin{center}
{{\bf Bernard de Wit}\\
Institute for Theoretical Physics \,\&\, Spinoza Institute,\\  
Utrecht University, The Netherlands\\
{\tt b.dewit@phys.uu.nl}}
\end{center}

\vskip 10mm

\begin{center} {\bf Abstract } \end{center}
\begin{quotation}\noindent
  Recent insights from string theory and supergravity on the
  macroscopic and the microscopic description of black hole entropy
  are discussed. \\
  \\[3mm]
\end{quotation}
\end{titlepage}
\eject
\section{Introduction}
It is a great pleasure and an honor to be a speaker at this symposium
on the occasion of Julius Wess' seventieth birthday.  Julius'
scientific achievements are impressive and his contributions to
supersymmetry have inspired the scientific activities of many of us
during a considerable part of our careers. In addition to his research
Julius has served the physics community in many other capacities.  Few
physicists of his stature have had such an impact on the structuring
of science and scientific education in and outside Germany. Here I am
also thinking of his more recent Balkan initiative to rebuild the
science infrastructure in former Yugoslavia after the disruptive and
painful events in the nineties.

On this occasion I would like to review certain aspects of the relation
between the macroscopic and microscopic descriptions of black holes.
Black holes are, roughly speaking, solutions of Einstein's equations
of general relativity that exhibit an event horizon. From inside this
horizon, nothing (and in particular, no light) can escape. In the
context of this talk, it suffices to think of spherically symmetric
and static black holes, with a flat space-time geometry at spatial
infinity. By definition, the region inside the horizon is not in the
backward lightcone of future timelike infinity. However, since the
discovery of Hawking radiation it has become clear that many of the
above classical features of black holes are modified.

I will be considering black holes in four space-time dimensions,
carrying electric and/or magnetic charges. Such solutions can be
described by Einstein-Maxwell theory, the classical field theory for
gravity and electromagnetism. The most general static black holes of
this type correspond to the Reissner-Nordstrom solutions,
characterized by a charge $Q$ and a mass $M$. In the presence of
magnetic charges, $Q$ is replaced by $\sqrt{q^2+p^2}$ in most
formulae, where $q$ and $p$ denote the electric and the magnetic
charge, respectively. For zero charges one is dealing with
Schwarzschild black holes. Two quantities associated with the black
hole horizon are the area $A$ and the surface gravity $\kappa_{\rm
  s}$. The area is simply the area of the two-sphere defined by the
horizon. The surface gravity, which is constant on the horizon, is
related to the force (measured at spatial infinity) that holds a unit
test mass in place. The mass $M$ and charge $Q$ of the black hole are
not directly associated with the horizon and can be expressed by
appropriate surface integrals at spatial infinity.

As is well known, there exists a striking correspondence between the
laws of thermodynamics and the laws of black hole mechanics
\cite{BCH}. Of particular importance is the first law, which, for
thermodynamics, states that the variation of the total energy is equal
to the temperature times the variation of the entropy, modulo work
terms, for instance proportional to a change of the volume. The
corresponding formula for black holes expresses how the variation of
the black hole mass is related to the variation of the horizon area,
up to work terms proportional to the variation of the angular
momentum. In addition there can also be a term proportional to a
variation of the charge, multiplied by the electric/magnetic potential
$\mu$ at the horizon. Specifically, the first law of thermodynamics,
$\delta E = T\,\delta S - p\,\delta V$, translates into
\begin{equation}
\label{first-law} 
\delta M = \frac{\kappa_{\rm s}}{2\pi} \, \frac {\delta A}{4}
+\mu\,\delta Q +\Omega\,\delta J\,. 
\end{equation} 
The reason for factorizing the first term on the right-hand side in
this particular form, is
that $\kappa_{\rm s}/2\pi$ represents precisely the Hawking
temperature \cite{Hawking}. This then leads to the identification of
the black hole entropy in terms of the horizon area, 
\begin{equation}
\label{area-law}
{\cal S}_{\rm macro} = \tfrac14 A\,,
\end{equation}
a result that is known as the area law \cite{Bek}.
In these equations the various quantities have been defined in Planck
units, meaning that they have 
been made dimensionless by multiplication with an appropriate power of
Newton's constant (we will set $\hbar=c=1$). This constant appears in the
Einstein-Hilbert  
Lagrangian according to ${\cal L}_{\rm EH} = -(16\pi\,G_{\rm N})^{-1}\,
\sqrt{\vert g\vert} \,R$. With this normalization the quantities
appearing in the first law are independent of the scale of the
metric. 

Einstein-Maxwell theory can be naturally embedded
into $N=2$ supergravity which may lead to an extension with a 
variety of abelian gauge fields and a related number of massless 
scalar fields (often called `moduli' fields, for reasons that will 
become clear later on). At spatial infinity these moduli fields will
tend to a constant, and the black hole mass will depend on these
constants, thus introducing additional terms on the right-hand side of
(\ref{first-law}).  

For Schwarzschild black holes the only relevant parameter is the mass
$M$ and I note the relations, 
\begin{equation}
\label{schwarzschild}
 A= 16\pi\,M^2\,,\qquad \kappa_{\rm s}= \frac{1}{4\,M}\;,
\end{equation}
consistent with (\ref{first-law}). For the 
Reissner-Nordstrom black hole, the situation is more subtle. Here one
distinguishes three different cases. For $M>Q$ one has the
non-extremal solutions, which exhibit two horizons, an exterior event
horizon and an interior 
horizon. When $M=Q$ one is dealing with an extremal black hole, for
which the two horizons coincide and the surface gravity vanishes. In
that case one has 
\begin{equation}
\label{e-RN}
 A= 4\pi\,M^2\,,\qquad \kappa_{\rm s}=0\,,\qquad \mu = Q\,
 \sqrt{\frac{4\pi}A}\,.  
\end{equation}
It is straightforward to verify that this result is consistent with
(\ref{first-law}) for variations in the subspace of extremal black
holes ({\it i.e.}, with $\delta M=\delta Q$). Because the surface
gravity vanishes, one might expect the entropy to vanish as suggested
by the third law of thermodynamics. Obviously, that is not the case as
the horizon area remains finite for zero surface gravity.  Finally,
solutions with $M<Q$ are not regarded as physically acceptable. Their
total energy is less than the electromagnetic energy alone and they no
longer have an event horizon but exhibit a naked singularity. Hence
extremal black holes saturate the bound $M\geq Q$ for physically
acceptable black hole solutions.

When embedding Einstein-Maxwell theory into a complete supergravity
theory, the above classification has an interpretation in terms of the
supersymmetry algebra. This algebra has a central extension
proportional to the black hole charge(s). Unitary representations of
the supersymmetry algebra must necessarily have masses that are larger
than or equal to the charge. When this bound is saturated, one is
dealing with so-called BPS supermultiplets. Such supermultiplets are
smaller than the generic massive $N=2$ supermultiplets and have a different
spin content. Because of this, BPS states are stable under (adiabatic)
changes of the coupling constants, and the relation between charge
and mass remains preserved. This important feature of BPS states will
be relevant for what follows.

\section{On macroscopic and microscopic descriptions}
\label{sec:macro-micro}
A central question in black hole physics concerns the
statistical interpretation of the black hole entropy. String theory
has provided new insights here \cite{Strominger:1996sh}, which have
led to important results. In this context it is relevant that strings
live in more that four space-time dimensions. In most situations the
extra dimensions are compactified on some internal manifold $X$ and
one is dealing with the usual Kaluza-Klein scenario leading to
effective field theories in four dimensions, describing low-mass modes
of the fields associated with certain eigenfunctions on the internal
manifold.

Hence the original space-time will locally be a product $M^4\times X$,
where $M^4$ denotes the four-dimensional space-time that one
experiences in daily life. I will denote the coordinates of $M^4$ by
$x^\mu$ and those of $X$ by $y^m$. In the situation described above
there exists a corresponding space $X$ at every point $x^\mu$ of
$M^4$, whose size is such that it will not be directly observable.
However, this space $X$ does not have to be the same at every point in
$M^4$, and moving through $M^4$ one may encounter various spaces $X$
which may or may not be equivalent.  Usually these spaces belong to
some well-defined class of fixed topology parametrized by certain
moduli. These moduli will appear as fields in the four-dimensional
effective field theory.  For instance, suppose that the spaces $X$ are
$n$-dimensional tori $T^n$. The metric of $T^n$ will appear as a field
in the four-dimensional theory and is related to the torus moduli.
Hence, when dealing with a solution of the four-dimensional theory
that is not constant in $M^4$, each patch in $M^4$ has a nontrivial
image in the space of moduli that parametrize the internal spaces $X$.

For a black hole solution, viewed in this higher-dimensional
perspective, the fields, and in particular the four-dimensional
space-time metric, will vary nontrivially over $M^4$, and so will the
internal space $X$.  When moving to the center of the black hole the
gravitational fields will become strong and the local product
structure into $M^4\times X$ could break down.  One feature of string
theory is, however, absent in this field-theoretic approach which
captures the local degrees of freedom of strings and branes: extended
objects carry also global degrees of freedom as they can wrap
themselves around nontrivial cycles of the internal space $X$. This
wrapping tends to take place at a particular position in $M^4$ so in
the context of the four-dimensional effective field theory, this will
reflect itself as a pointlike object. This wrapped object is the
string theory representation of the black hole!

One is thus dealing with two complementary pictures of the black hole.
One based on general relativity where a point mass generates a global
solution of space-time with strongly varying gravitational fields,
which I shall refer to as the {\it macroscopic} description.  The
other one, based on the internal space where an extended object is
entangled in one of its cycles, does not involve gravitational fields
and can easily be described in flat space-time.  This description will
be refered to as {\it microscopic}. To understand how these two
descriptions are related is far from easy, but a connection must exist
in view of the fact that gravitons are closed string states which
interact with the wrapped branes. These interactions are governed by
the string coupling constant $g_{\rm s}$ and one is thus confronted
with an interpolation in that coupling constant. In principle, such an
interpolation is very difficult to carry out, so that a realistic
comparison between microscopic and macroscopic results is usually
impossible. However, reliable predictions are possible for extremal
black holes! In a supersymmetric setting extremal black holes are BPS
and, as I indicated earlier, in that situation there are reasons to
trust such interpolations. Indeed, it has be shown that the
predictions based on these two alternative descriptions can be
successfully compared and new insights about black holes can be
obtained.

But how do the wrapped strings and branes represent themselves in the
effective action description and what governs their interactions with
the low-mass fields? Here it is important to realize that the massless
four-dimensional fields are associated with harmonic forms on $X$.
Harmonic forms are in one-to-one correspondence with so-called
cohomology groups consisting of equivalence classes of forms that are
closed but not exact. The number of independent harmonic forms of a
given degree is given by the so-called Betti numbers, which are fixed
by the topology of the spaces $X$. When expanding fields in a
Kaluza-Klein scenario, the number of corresponding massless fields can
be deduced from an expansion in terms of tensors on $X$
corresponding to the various harmonic forms. The higher-dimensional
fields $\Phi(x,y)$ thus decompose into the massless fields $\phi^A(x)$
according to (schematically),
\begin{equation}
\label{KK-decomposition}
\Phi(x,y) = \phi^A(x)\;\omega_A(y)\,,
\end{equation}
where $\omega_A(y)$ denotes the independent harmonic forms on $X$. The above
expression, when substituted into the action of the higher-dimensional
theory, leads to interactions of the fields $\phi^A$ proportional to
the `coupling constants', 
\begin{equation}
\label{interaction}
C_{ABC\cdots} \propto \int_X \omega_A\wedge\omega_B\wedge\omega_C
\cdots\,. 
\end{equation}
These constants are known as intersection numbers, for reasons that
will become clear shortly. 

I already mentioned that the Betti numbers depend on the topology of
$X$. This is related to Poincar\'e duality, according to which
cohomology classes are related to homology classes. The latter consist
of submanifolds of $X$ without boundary that are themselves not a
boundary of some other submanifold of $X$. This is precisely relevant
for wrapped branes which indeed cover submanifolds of $X$, but are not
themselves the boundary of a submanifold because otherwise the brane
could collapse to a point. Without going into detail, this implies
that there exists a dual relationship between harmonic $p$-forms
$\omega$ and $(d_X-p)$-cycles, where $d_X$ denotes the dimension of
$X$. One can therefore choose a homology basis for the $(d_X-p)$-cycles
dual to the basis adopted for the $p$-forms.  Denoting this basis by
$\Omega_A$ the wrapping of an extended object can now be characterized
by specifying its corresponding cycle ${\cal P}$ in terms of the homology
basis,
\begin{equation}
\label{intersection}
{\cal P} = p^A\,\Omega_A\,.
\end{equation}
The integers $p^A$ count how many times the extended object is wrapped
around the corresponding cycle, so one is actually dealing with
integer-valued cohomology and homology. The wrapping numbers $p^A$
reflect themselves as magnetic charges in the effective action. The
electric charges are already an integer part of the effective action,
because they are associated with gauge transformations that usually
originate from the higher-dimensional theory. 

Owing to Poincar\'e duality it is thus very natural that the winding
numbers interact with the massless modes in the form of magnetic
charges, so that they can be incorporated in the effective action.
Before closing this section, I note that, by Poincar\'e duality, one
can express the number of intersections by
\begin{equation}
\label{intersections} 
{\cal P}\cdot {\cal P}\cdot {\cal P}\cdots
=C_{ABC\cdots}\,p^Ap^Bp^C\cdots\;. 
\end{equation}
This is a topological characterization of the wrapping, which will
appear in later formulae.

\section{Black holes in M/String Theory}
\label{sec:M-bh} 
As an example I now discuss the black hole entropy derived from both
microscopic and macroscopic arguments in a special case.  Consider
M-theory, which, in the strong coupling limit of type-IIA string 
theory, is described by eleven-dimensional supergravity. The latter is
invariant under 32 supersymmetries. Seven of the eleven space-time
dimensions are compactified on an internal space which is the product
of a Calabi-Yau threefold (a three-dimensional complex manifold, which
henceforth I denote by $CY_3$) times a circle $S^1$. Such a space
breaks part of the supersymmetries and only 8 of them are left
unaffected. In the context of the four-dimensional space-time $M^4$,
these 8 supersymmetries are encoded into two independent Lorentz
spinors and for that reason this symmetry is referred to as $N=2$
supersymmetry. Hence the effective four-dimensional field theory will
be some version of $N=2$ supergravity.

M-theory contains a five-brane and this is the microscopic object that is
responsible for the black holes that I consider; the five-brane has
wrapped itself on a 4-cycle ${\cal P}$ of the ${CY}_3$ space
\cite{Maldacena:1997de}.  Alternatively one may consider this class of
black holes in type-IIA string theory, with a 4-brane wrapping the
4-cycle \cite{Vafa:1997gr}. The 4-cycle is subject to certain
requirements which will be mentioned in due course.

The massless modes captured by the effective field theory correspond
to harmonic forms on the $CY_3$ space; they do not depend on the $S^1$
coordinate. The 2-forms are of particular interest. In the effective
theory they give rise to vector gauge fields $A_\mu{}^A$, which
originate from the rank-three tensor gauge field in eleven dimensions.
In addition there is an extra vector field $A_\mu{}^0$ coresponding to
a 0-form which is related to the graviphoton associated with $S^1$.
This field will couple to the electric charge $q_0$ associated with
momentum modes on $S^1$ in the standard Kaluza-Klein fashion. The
2-forms are dual to 4-cyles and the wrapping of the five-brane is
encoded in terms of the wrapping numbers $p^A$, which appear in the
effective field theory as magnetic charges which couple to the gauge
fields $A_\mu{}^A$. Here one sees Poincar\'e duality at work, as the
magnetic charges couple nicely to the corresponding gauge fields. For
a Calabi-Yau three-fold, there is a triple intersection number
$C_{ABC}$ which appears in the three-point couplings of the effective
field theory. There is a subtle topological feature that I have not
explained before, which is typical for complex manifolds containing
4-cycles, namely the existence of another quantity of topological
interest known as the second Chern class. The second Chern class is a
4-form whose integral over a four-dimensional Euclidean space defines
the instanton number. The 4-form can be integrated over the 4-cycle
${\cal P}$ and yields $c_{2A} \,p^A$, where the $c_{2A}$ are integers.

Let me now turn to the microscopic counting of degrees of freedom
\cite{Maldacena:1997de}.  These degrees of freedom are associated with
the massless excitations of the wrapped five-brane characterized by
the wrapping numbers $p^A$ on the 4-cycle. The 4-cycle ${\cal P}$ must
correspond to a holomorphically embedded complex submanifold in order
to preserve 4 supersymmetries.  The massless excitations of the
five-brane are then described by a $(1+1)$-dimensional superconformal
field theory (the reader may also consult \cite{MinMooTsi}).  Because
I have compactified the spatial dimension on $S^1$, one is dealing
with a closed string with left- and right-moving states. The 4
supersymmetries of the conformal field theory reside in one of these
two sectors, say the right-handed one.  Conformal theories in $1+1$
dimensions are characterized by a central charge, and in this case
there is a central charge for the right- and for the left-moving sector
separately. These central charges are expressible in terms of the
wrapping numbers $p^A$ and depend on the intersection numbers and the
second Chern class, according to
\begin{eqnarray}
  \label{eq:c-LR}
  c_L &=& C_{ABC}\, p^Ap^Bp^C + c_{2A}\,p^A \,,\nonumber \\
  c_R &=& C_{ABC}\, p^Ap^Bp^C + \tfrac12  c_{2A}\,p^A \,.
\end{eqnarray}
I should stress that the above result is far from obvious and holds
only under the condition that the $p^A$ are large. 
In that case every generic deformation of ${\cal P}$ will be smooth.
Under these circumstances it is possible to relate the topological
properties of the 4-cycle to the topological data of the Calabi-Yau
space. 

One can now choose a state of given momentum $q_0$ which is supersymmetric
in the right-moving sector. From rather general arguments it follows that
such states exist. The corresponding states in the left-moving sector
have no bearing on the supersymmetry and these states have a certain
degeneracy depending on the value of $q_0$. In this way one has a
tower of BPS states invariant under 4 supersymmetries, built on
supersymmetric states in the right-moving sector and comprising
corresponding degenerate states in the left-moving sector. One can
then use Cardy's formula, which states that the degeneracy of states
for fixed but large momentum (large as compared to $c_L$) equals
$\exp[2\pi\sqrt{\vert q_0\vert \, c_L/6}]$.  This leads to the
following expression for the entropy,
\begin{equation}
  \label{eq:S-CY-micro}
  {\cal S}_{\rm micro}(p,q) = 2\pi\sqrt{ \tfrac16 
  \vert\hat q_0\vert (C_{ABC} \,p^Ap^Bp^C + c_{2A}\,p^A)}\,,
\end{equation}
where $q_0$ has been shifted according to $\hat q_0 = q_0 + \tfrac1{2}
C^{AB} q_Aq_B$, with $C^{AB}$ the inverse of $C_{AB}= C_{ABC}p^C$.
This modification is related to the fact that the electric charges
associated with the gauge fields $A_\mu{}^A$ will interact with the
M-theory two-brane \cite{Maldacena:1997de}. The existence of this
interaction can be inferred from the fact that the two-brane interacts
with the rank-three tensor field in eleven dimensions, from which the
vector gauge fields $A_\mu{}^A$ originate.

I stress that the above results apply in the case of large charges.
The first term proportional to the triple intersection number is
obviously the leading contribution whereas the terms proportional to the
second Chern class are subleading. The importance of the subleading
terms will become more clear in later sections. Having obtained a
microscopic representation of a BPS black hole, it now remains to make
contact with it by deriving the corresponding black hole solution
directly in the $N=2$ supergravity theory. This is discussed below. 

\section{An entropy formula for $N=2$ supergravity}
\label{sec:entropy-formula}
The charged black hole solutions in $N=2$ supergravity are invariant
under 4 of the 8 supersymmetries. They are solitonic, and interpolate
between fully supersymmetric configurations at the horizon and at
spatial infinity. At spatial infinity, where the effect of the charges
can be ignored, one has flat Minkowski space-time. The scalar moduli
fields tend to certain (arbitrary) values on which the black hole mass
will depend.  At the horizon the situation is rather
different, because here the charges are felt and one is not longer
dealing with flat space-time, but with a so-called Bertotti-Robinson
space, ${\rm AdS}_2\times S^2$.  In that situation the requirement of
full $N=2$ supersymmetry is highly restrictive and for spherical
geometries one can prove that the values of the moduli fields at the
horizon are in fact fixed in terms of the charges. The corresponding
equations are known as the attractor equations
\cite{Ferrara:1995ih,Strominger:1996kf, Ferrara:1996dd} and they apply
quite generally. For effective actions with interactions quadratic in
the curvature the validity of these attractor equations was
established in \cite{LopesCardoso:2000qm}.

The attractor equations play a crucial role as they ensure that the
entropy, a quantity that is associated with the horizon, will depend
on the black hole charges and not on other quantities, in line with
the microscopic results presented in the previous section. Hence one
is interested in studying charged black hole solutions which are BPS,
meaning that they are invariant under 4 supersymmetries. The matter
supermultiplets contain the gauge fields $A_\mu{}^A$ coupling to
electric and magnetic charges $q_A$ and $p^A$, respectively. In
addition there is one extra graviphoton field $A_\mu{}^0$ which may
couple to charges $q_0$ and $p^0$. When comparing to the solutions of
the previous section, one obviously sets $p^0=0$, although from the
supergravity point of view there is no need for such a restriction.

However, there is an infinite variety of $N=2$ supergravity actions
coupling to vector multiplets. Fortunately these actions can be
conveniently encoded into holomorphic functions that are homeogeneous
of second degree \cite{DWVHVPL}. In the case at hand the simplest
action is, for instance, based on the function,
\begin{equation}
  \label{eq:CY-sg}
  F(Y)=-\frac1{6}  \frac{C_{ABC}\,Y^AY^BY^C}{Y^0} \,,
\end{equation}
where the holomophic variables $Y^I$ ($I=0,A$) are associated with the
vector multiplets; they can be identified projectively with the scalar
moduli fields that are related to a subset of the moduli of the
Calabi-Yau space.  The black hole solution will thus encode the changes in the 
Calabi-Yau manifold when moving from the black hole horizon
towards spatial infinity, precisely as discussed in a more general
context in section~2. Note the presence of the triple intersection
form $C_{ABC}$ which will appear in the interaction vertices of the
corresponding Lagrangian. 

The attractor equations also involve the function $F(Y)$. In terms of
the quantities $Y^I$ and the first derivatives of $F(Y)$, they take
the form,
\begin{equation}
  \label{eq:attractor}
  Y^I-\bar Y^I = \mathrm{i} p^I\,,\qquad 
  F_I(Y) - \bar F_I(\bar Y) = \mathrm{i} q_I\,,
\end{equation}
where $F_I(Y)= \partial F(Y)/\partial Y^I$. In principle these
equations yield the horizon values of the $Y^I$ in terms of the
charges. Depending on the values of the charges and on the complexity
of the function $F(Y)$, it may not be possible to write down
solutions in closed form.

The action corresponding to (\ref{eq:CY-sg}) gives rise to a black
hole solution with charges $p^A$, $q_A$ and $q_0$ (with $p^0=0$). Its
area can be calculated and is equal to
\begin{equation}
  \label{eq:S-CY-macro}
  A(p,q) = 8 \pi\sqrt{ \tfrac16 
  \vert\hat q_0\vert \,C_{ABC} \,p^Ap^Bp^C }\,.
\end{equation}
Upon invoking the area law this result leads precisely to the first
part of the microscopic entropy (\ref{eq:S-CY-micro}). One has thus
reproduced the leading contributions to the entropy from supergravity.

How can one reproduce the subleading terms in view of the fact that
these terms scale differently whereas the function $F(Y)$ and the
attractor equations all seem to scale uniformly? To explain how this
is resolved I must first spend a few words on the reason why the
function $F(Y)$ was homogeneous in the first place. The covariant
fields corresponding to a vector supermultiplets comprise a so-called
restricted chiral multiplet, which can be assigned a unique (complex)
scaling weight. The $Y^I$ are proportional to the lowest component of
these multiplets, and can be assigned the same scaling weight. Any
(holomorphic) function of these restricted multiplets will define a
chiral superfield, whose chiral superspace integral will lead to a
supersymmetric action.  However, in order to be able to couple to
supergravity, this function must be homogeneous of second degree
\cite{DWVHVPL}.  To deviate from this homogeneity pattern in the
determination of the entropy, one needs to introduce a new type of
chiral superfield whose value at the horizon will be fixed in a way
that breaks the uniformity of the scaling.  There exists such a
multiplet.  Namely, from the fields of (conformal) supergravity
itself, one can again extract a restricted chiral multiplet, which in
this case comprises the covariant quantities associated with the
supergravity fields.  This time the restricted chiral multiplet is not
a scalar, but an auxiliary anti-selfdual tensor, and just as before it
can be assigned a unique scaling weight. Its lowest component is an
auxiliary field that is often called the graviphoton field strength,
which is strictly speaking a misnomer because it never satisfies a
Bianchi identity. It appears in the transformation rule of the
gravitino fields and in simple Lagrangians its field equations
express it in terms of moduli-dependent linear combinations of other
vector field strengths.  The square of this restricted tensor defines
a complex scalar field which constitutes the lowest component of a
chiral supermultiplet and which is proportional to a field I will
denote by $\Upsilon$, such that the (complex) scaling weight that can be
assigned to $\Upsilon$ is twice that of the $Y^I$. More general
supergravity Lagrangians can then described by holomorphic functions
$F(Y,\Upsilon)$ that are homogeneous of degree 2, {\it i.e.},
\begin{equation}
  \label{eq:homogeneous-F}
  F(\lambda Y^I,\lambda^2 \Upsilon) = \lambda^2\, F( Y^I,\Upsilon)\,.
\end{equation}
A nontrivial dependence on $\Upsilon$ in the function $F(Y,\Upsilon)$
has important consequences, because the supermultiplet of which
$\Upsilon$ is the first component contains other components with
terms quadratic in the Riemann tensor.  Hence actions based on a
function (\ref{eq:homogeneous-F}) with a nontrivial dependence on
$\Upsilon$ will contain terms proportional to the square of the
Riemann tensor, multiplied by the first derivative of $F$ with respect
to $\Upsilon$.  The attractor equations (\ref{eq:attractor}) remain
valid with $F(Y)$ replaced by $F(Y,\Upsilon)$. However, the field
$\Upsilon$ has its own independent attractor value; at the horizon it 
must be equal to $\Upsilon=-64$, independent of the charges.  This
phenomenon explains why the area or the entropy formula are not
necessarily a homogeneous function of the charges.

To fully reproduce the entropy formula (\ref{eq:S-CY-micro}) including
the subleading terms proportional to the second Chern class, one may
attempt a Lagrangian based on the function
\begin{equation}
  \label{eq:CY+cc-sg}
  F(Y)=-\frac1{6} \, \frac{C_{ABC}\,Y^AY^BY^C}{Y^0} - \frac{c_{2A}} 
   {24\cdot 64}\, \frac{ Y^A}{Y^0} \,\Upsilon \,,
\end{equation}
which is indeed holomorphic and homogeneous. On the basis of this
modification one can again calculate the horizon area in the hope of
recovering the entropy (\ref{eq:S-CY-macro}) upon use of the area law.
However, the result is negative and it seems obvious that no solution
can be found in this way \cite{BCDWLMS}. 

At this point the only way out is to no longer rely on the area law in
extracting a value for the entropy. Indeed the area law is not
expected to hold for actions that supersede the Einstein-Hilbert one.
However, Wald has proposed an alternative definition of black hole
entropy which can be used for any Lagrangian that is invariant under
general coordinate transformations, and which is based on the
existence of a conserved surface charge \cite{WaldIyer}. The latter is
related to the conventional Noether current associated with general
coordinate transformations, which, for a gauge symmetry, can be
written as a pure improvement term: the divergence of an antisymmetric
tensor, called the Noether potential. It turns out that with the help
of the Noether potential one can define a surface charge integrated
over the boundary of a Cauchy surface, which for the black hole
extends from spatial infinity down to the horizon. Changes in the
continuous variety of black hole solutions should leave this charge 
unchanged.  Under certain conditions one can show that the change of
the surface integral at spatial infinity corresponds to the mass and
angular momentum variations in the first law. Therefore one identifies
the surface integral at the horizon with the entropy, so that the
validity of the first law will remain ensured.

I should stress that there are various subtle points here, some of which 
have been discussed in \cite{CarDeWMoh}.  The prescription based
on the surface charge can be applied to standard Einstein gravity, in
which case one just recovers the area law. But for theories with
higher-derivatives, there are nontrivial correction terms, which
follow from a calculation of the Noether potential. I should caution
the reader that the relevant correction term in the case at hand does
in fact not reside in the terms quadratic in the Riemann tensor, but
in some other terms related to them by supersymmetry.

From an evaluation of the Noether potential, taking into account all
the constraints imposed by the supersymmetry at the horizon, it 
follows that the entropy can be written in a universal form
\cite{LopesCardoso:1998wt},
\begin{equation}
  \label{eq:W-entropy}
  {\cal S}_{\rm macro} (p,q) = \pi \Big[\vert Z\vert^2 - 265\, 
  {\rm Im} F_\Upsilon \Big ]_{\Upsilon=-64} \,. 
\end{equation}
Here the first term denotes the Bekenstein-Hawking entropy, because
$\vert Z\vert^2= p^IF_I(Y,\Upsilon)- q_IY^I$ is just the area in
Planck units divided by $4\pi$. This term is clearly affected by the
presence of the higher-order derivative interactions. On top of that
there is a second term proportional to the derivative of
$F(Y,\Upsilon)$ with respect to $\Upsilon$. This term
thus represents the deviation of the area law. The above formula applies
to any $N=2$ supergravity solution.

Because of the fact that all quantities of interest are directly
related to the holomorphic and homogeneous functions
(\ref{eq:homogeneous-F}), the determination of the area and entropy is
merely an algebraic exercise, which no longer requires to construct
the full solution. Given the function $F(Y,\Upsilon)$ one first
attempts to solve the attractor equations (\ref{eq:attractor}) with
$F_I(Y)$ replaced by $F_I(Y,\Upsilon)$. Subsequently one determines
the area and the entropy in terms of the charges. For the function
(\ref{eq:CY+cc-sg}) this was shown to lead precisely to the
microscopic entropy formula (\ref{eq:S-CY-micro}). To exhibit the
deviation from the area law, I also give the area (which is obviously
not known from microscopic considerations),
\begin{equation}
  \label{eq:area/entropy}
 \tfrac1{4} A(p,q) = \frac{C_{ABC}\, p^Ap^Bp^C + \tfrac12 c_{2A}\,p^A}
  {C_{ABC}\, p^Ap^Bp^C + c_{2A}\,p^A} \;
 {\cal S}_{\rm macro} (p,q) \,. 
\end{equation}
Interestingly the proportionality factor is just the ratio
$c_R/c_L$ of the two central charges defined in (\ref{eq:c-LR}). By
combining new ingredients from supergravity and general relativity it
is thus possible to fully account for the black hole entropy that is
obtained by counting microstates.

Before moving to the next section I wish to add some observations. In
addition to being able to evaluate the properties of the black hole
solution at the horizon one should also want to understand the full
structure of the BPS black holes away from the horizon, in the
presence of the interactions quadratic in the Riemann curvature. This
was the subject of \cite{LopesCardoso:2000qm}, where a rather general
class of such solutions was studied, including multi-centered ones.
I refer to that work for further details. It is also worth
pointing out that sofar I have been basing myselves on the 
effective Wilsonian action. A priori, one does
not expect that the final macroscopic description of black hole
mechanics can be obtained within a Wilsonian framework.

\section{Heterotic black holes}
\label{sec:heter-black-holes}
In \cite{CarDeWMoh,LopesCardoso:1999ur} the modified entropy formula
(\ref{eq:W-entropy}) was applied to heterotic black holes. Although
the formula is derived for $N=2$ supergravity, the result can readily
be generalized to the case of heterotic $N=4$ supersymmetric
compactifications. This involves an extension of the target-space
duality group to ${\rm SO}(6,22)$ with a corresponding extension to 28
electric and 28 magnetic charges that take their values in a
$\Gamma^{6,22}$ lattice. The $N=4$ supersymmetric heterotic models
have dual realizations as type-II string compactifications on
$K3\times T^2$. In contrast to $N=2$ Calabi-Yau compactifications, the
holomorphic function which encodes the effective Wilsonian action is
severely restricted in the $N=4$ case.  Therefore it is often possible
to obtain exact predictions in this context.

The relevant function for the heterotic case takes the following
form in lowest order,
\begin{equation}
  \label{eq:F-het}
  F(Y) = - \frac{Y^1\,Y^a\eta_{ab}Y^b}{Y^0} \,,
\end{equation}
where $a,b= 2,\ldots,n$. This function will be modified in due course
by a function of $\Upsilon$ and of the dilaton field $S=-\mathrm{i}
Y^1/Y^0$.  Let me first consider (\ref{eq:F-het}) in the absence of
these modifications.  Then the $2n$ scalar moduli in the effective
action are described by a nonlinear sigma model with the following
target space,
\begin{equation}
  \label{eq:heterotic-duality}
  {\cal M} = \frac{{\rm SU}(1,1)}{{\rm U}(1)} \times \frac{{\rm SO}(2,n-1)}
   {{\rm SO}(2)\times {\rm SO}(n-1)} \,. 
\end{equation}
The electric and magnetic charges transform under the action of the
${\rm SU}(1,1)\times{\rm SO}(2,n-1)$ isometry group. The first factor,
${\rm SU}(1,1)$, is associated with $S$-duality. This is a strong-weak
coupling duality which interchanges electric and magnetic charges. The
second factor, ${\rm SO}(2,n-1)$, is associated with $T$-duality (also
called target-space duality). There is a technical complication in the
description based on (\ref{eq:F-het}) because the charges that follow
from the $Y^I$ through the attractor equations are not in a convenient
basis for $S$- and $T$-duality. A proper basis is found upon
interchanging the electric and magnetic charges $q_1$ and $p^1$ by
an electric/magnetic duality. Fortunately there is no need to discuss
this in any detail, as the entropy and area of the corresponding black
holes depend only on $T$-duality invariants of the charges. Note that
in the extension to $N=4$ the second factor of the target space
(\ref{eq:heterotic-duality}) changes into ${\rm SO}(6,22) /[{\rm 
  SO}(6)\times {\rm SO}(22)]$; this space is parametrized by the 132
scalar fields belonging to 22 $N=4$ vector supermultiplets.

To be specific let me first present the result for the entropy and
horizon area for the solution based on (\ref{eq:F-het}) as a function
of the charges,
\begin{equation}
  \label{eq:class-het-entropy}
  {\cal S}_{\rm macro}(p,q)= \tfrac14 A(p,q)  
  = \pi \sqrt{q^2\, p^2 -(p\cdot q)^2} \,.
\end{equation}
Here I used $T$-duality invariant combinations of the charges, defined by 
\begin{eqnarray}
  \label{eq:charge-invariants}
  q^2&=& 2\,q_0p^1 - \tfrac12 q_a\eta^{ab} q_b\,, \nonumber \\
  p^2 &=& -2 p^0q_1 - 2\,p^a\eta_{ab}p^b\,, \nonumber \\
  p\!\cdot \!q &=& q_0p^0-q_1p^1 + q_ap^a \,,
\end{eqnarray}
where the $p^I$ and $q_I$ on the right-hand side are the charges that
appear in the attractor equations (\ref{eq:attractor}) based on the
function (\ref{eq:F-het}). While the combinations
(\ref{eq:charge-invariants}) are invariant under $T$-duality, they
transform as an ${\rm SO}(2,1)$ vector under $S$-duality, such that
the entropy formula (\ref{eq:class-het-entropy}) is invariant under
both $T$- and $S$-duality.  The $S$-duality transformations, which
constitute the group $\mathrm{SL}(2,\mathbb{Z})$, of the charges and
of the dilaton field are related through the attractor equations.

In the $N=4$ theory, the charge lattice is $S$-duality invariant,
meaning that the above transformations always leads to another point
on the lattice that is physically realized. Note that the supergravity
calculations yield no intrinsic definition of the normalization of the
charge lattice and consequently the dilaton normalization is a priori
not known. Hence the precise characterization of the arithmetic
subgroup of $\mathrm{SL}(2)$ that defines the $S$-duality group is not
obvious, but the crucial point is that the normalization of the
dilaton is related to the normalization of the lattice of charges.
Later on in this section I will relate the supergravity results to
microscopic data which will confirm the above identifications. Observe
that the invariants $p^2$ and $q^2$ are not positive definite. In fact
in the limit of large charges they will both become negative.

When adding a function to (\ref{eq:F-het}) proportional to $\Upsilon$
and depending otherwise on the dilaton field $S$, it turns out that
the target-space duality remains unaffected. However, $S$-duality is
affected in general so the question is whether there exists a
specific modification that leaves $S$-duality intact. This turns out
to be the case, but one is forced to accept a certain amount of
non-holomorphicity in the description \cite{LopesCardoso:1999ur}.  To
derive this is rather nontrivial and I simply quote the result in a
form that was indicated in \cite{CDWKM},
\begin{equation}
  \label{eq:F-het-mod}
  F(Y,\bar Y,\Upsilon,\bar \Upsilon) = - \frac{Y^1\,Y^a\eta_{ab}Y^b}{Y^0} + 
\frac{\mathrm{i}}{64\pi} 
   \,\Upsilon \log\eta^{12}(S)  +\frac{\mathrm{i}}{128\pi}\, 
  (\Upsilon+\bar \Upsilon)   \log (S+\bar S)^6 \,,
\end{equation}
where the non-holomorphic corrections reside in the last term.  Here
$\eta(S)$ denotes the Dedekind eta-function, which satisfies the
asymptotic formula $\log\eta(S)\approx -\frac1{12} \pi\,S - {\mathrm
  e}^{-2\pi S}+ {\cal O}({\mathrm e}^{-4\pi S})$ for large positive
real values of $S$. The presence of non-holomorphic terms in
(\ref{eq:F-het-mod}) is not entirely unexpected: the Wilsonian
couplings are holomorphic but do not necessarily reflect the
symmetries of the underlying theory, while the physical couplings must
reflect the symmetry and may thus have different analyticity
properties. The underlying reason is that the Wilsonian action is
based on integrating out the massive modes above a certain scale; the
effect of the massless modes must thus be included separately. The
massless modes give rise to a certain non-local action, which is
supersymmetric although it is not based on a chiral superspace
density. Its explicit form has not been given so far, so that the
non-holomorphic terms have been determined directly by requiring
$S$-duality of the attractor equations and the entropy, and
consistency with string perturbation theory. The results are in accord
with the $N=4$ results of \cite{Harvey:1996ir}. The normalization of
the $R^2$-terms in the effective action has been fixed by using
string-string duality, or, alternatively, by requiring agreement with
the known asymptotic degeneracy of electrically charged black holes.

Including the non-holomorphic corrections, the result of
\cite{LopesCardoso:1999ur} can be summarized as follows. The
non-trivial attractor equations are the ones that determine the
horizon value of the complex dilaton field $S$ in terms of the black
hole charges. They read as follows (we have now set $\Upsilon$ to its
horizon value),
\begin{eqnarray}
\label{eq:nonholostab}
\vert S\vert^2 \, p^2 &=& q^2 -  \frac{ 2}{\pi} ( S + {\bar S}) \, 
\Big(S \frac{\partial}{\partial S}+ \bar S \frac{\partial}{\partial \bar
  S}\Big) 
 \log\left[ (S+\bar S)^6\vert\eta(S)\vert^{24}\right] \;, \nonumber\\ 
(S - {\bar S}) \,p^2 &=& {}-2 \,\mathrm{i} \, p\!\cdot \!q + \frac{2}{\pi} 
 (S + {\bar S}) \, \Big(\frac{\partial}{\partial S} 
 - \frac{\partial}{\partial\bar S}\Big) 
\log\left[ (S+\bar S)^6\vert \eta(S)\vert^{24}\right] \;.
\end{eqnarray}
The expression for the macroscopic entropy reads, 
\begin{eqnarray}
  \mathcal{S}_\text{macro} = 
- \pi \left[ \frac{q^2 - \mathrm{i} p\!\cdot\! q \, 
(S - {\bar S}) + p^2 \,|S|^2} 
{S + {\bar S}} \right] -2 \, \log\left[ (S + {\bar S})^6
  |\eta(S)|^{24}\right] \;, 
\label{eq:nonholoentropy}
\end{eqnarray}  
with the dilaton field subject to (\ref{eq:nonholostab}). The first
term in this equation corresponds to one-fourth of the horizon area,
which, via (\ref{eq:nonholostab}), is affected by the various
corrections.  The second term represents an extra modification, which
explicitly contains the non-holomorphic correction. Both terms are
invariant under target-space duality and $S$-duality. As explained
above, $S$-duality was achieved at the price of including non-holomorphic
terms, here residing in the $\log(S+\bar S)$ terms.

In the $N=4$ setting one can distinguish two types of BPS-states.
Purely electric or magnetic configurations constitute 1/2-BPS states,
whereas dyonic ones are 1/4-BPS states. For $N=2$ the distinction
between the two types of states disappears and one has only 1/2-BPS
states. In the context of $N=4$, the generic BPS states are the dyonic
ones, characterized by a nonzero value for $q^2\,p^2
-(p\!\cdot\!q)^2$, while the 1/2-BPS states are subject to the
condition $q^2\,p^2-(p\!\cdot\!q)^2=0$.  In the remainder of this
section I restrict myself to the dyonic states. The macroscopic
results given above can be confronted with an explicit formula for the
microscopic degeneracy of BPS dyons in four-dimensional $N=4$ string
theory proposed in \cite{Dijkgraaf:1996it}. This proposal generalizes
the expression for the degeneracies of electric heterotic string
states (to be presented in the next section), to an expression that
depends on both electric and magnetic charges such that it is formally
covariant with respect to $S$-duality. In \cite{Dijkgraaf:1996it} it
was already shown that the dyonic degeneracy was consistent with the
area law, {\it i.e.}  with (\ref{eq:class-het-entropy}) in the limit
of large charges. Recently, this degeneracy was derived from the known
degeneracies of black holes in five spacetime dimensions
\cite{ShiStroYin}.

For my purpose it suffices to note that the degeneracy formula can be
expressed in terms of an integral over an appropriate 3-cycle that
involves an automorphic form $\Phi_{10}(\Omega)$,
\begin{equation}
  \label{eq:dvvdeg}
  d(q, p) = \oint {\mathrm d} \Omega \,
 \frac{{\mathrm e}^{\mathrm{i}\pi (Q^T \Omega \,Q)}}{\Phi_{10}(\Omega)} \;.
\end{equation}
Here $\Omega$ denotes the period matrix for a genus-2 Riemann surface,
which parametrizes the $\mathrm{Sp}(2)/\mathrm{U}(2)$ cosets; it 
can be written as a complex, symmetric, two-by-two matrix,
\begin{equation}
  \label{eq:oq}
  \Omega = 
  \begin{pmatrix} 
    \rho & \upsilon \\ \upsilon &\sigma 
  \end{pmatrix}\,.
\end{equation}
In the exponent of the numerator of (\ref{eq:dvvdeg}) the direct
product of the period matrix with the invariant metric of the charge
lattice is contracted with the charge vector comprising the 28
magnetic and 28 electric charges, so that $Q^T \Omega \,Q = \rho\, p^2
+ \sigma\,q^2 +2\, \upsilon\,p\!\cdot\!q$, where $p^2$, $q^2$ and
$p\!\cdot\! q$ were defined previously in
(\ref{eq:charge-invariants}).  The inverse of $\Phi_{10}$ has poles
and the formula~(\ref{eq:dvvdeg}) picks up a corresponding residue
whenever the 3-cycle encloses such a pole. However, the poles are
located in the interior of the Siegel half-space and not just at its
boundary and therefore the choice of the 3-cycles is rather subtle.

In \cite{CDWKM} the degeneracy formula was studied by means of a
saddle-point approximation in the limit of large charges, but now
retaining also the subleading terms. Remarkably enough the
result is in precise agreement with the macroscopic results, {\it
  i.e.}  (\ref{eq:nonholostab}) and (\ref{eq:nonholoentropy}),
including the non-holomorphic terms. The equations
(\ref{eq:nonholostab}) turn out to correspond to the equations that
determine the location of the saddle-point, while
(\ref{eq:nonholoentropy}) represents the value of the integrand in
(\ref{eq:dvvdeg}) taken at the saddle-point, including the results
from integrating out the fluctuations about the saddle-point. This
shows that the macroscopic entropy, defined by
(\ref{eq:nonholoentropy}) as a function of the charges and the dilaton
field, is in fact stationary under variations of the latter. 

\section{The area law and elementary string states}
\label{sec:area}
The area law is clearly violated in the presence of the subleading
corrections, as is shown in the $N=2$ entropy
formula~(\ref{eq:W-entropy}). Of course, it depends on the theory in
question and on the values for the charges, how sizable this violation
is.  A particularly interesting case emerges for black holes for which
the leading contribution to the entropy and area vanish. In that case,
the subleading terms become dominant and (\ref{eq:area/entropy}) shows
that the area law is replaced by ${\cal S}_{\text{macro}}= \tfrac12
A(p,q)$, whereas the typical dependence on the charges proportional to
the square root of a quartic polynomial is changed into the square
root of a quadratic polynomial. It is easy to see how this can be
accomplished for the heterotic black holes, namely, by suppressing all
the charges $p^0,q_1,p^2,\ldots,p^n$ in (\ref{eq:charge-invariants})
leaving $q^2$ as the only nonvanishing $T$-duality invariant charge
combination.  This is a remarkable result. In fact these states are
precisely generated by perturbative heterotic string states arising
from a compactification of six dimensions. In the supersymmetric
right-moving sector they carry only momentum and winding and contain
no oscillations, whereas in the left-moving sector oscillations are
subject to the string matching condition.  The oscillator number is
then linearly related to $q^2$. These string states are 1/2-BPS states
and correspond to electrically charged states (possibly upon a
suitable electric/magnetic duality redefinition).

Precisely these perturbative states already received quite some attention
in the past (for an early reference, see \cite{DaHa}). Because the
higher-mass string BPS states are expected to be within their
Schwarzschild radius, it was conjectured that they should have an
interpretation as black holes. The idea that elementary particles, or
string states, are behaving like black holes, is not new and has been
around for quite some time. 
Hence, their calculable level density, proportional to the exponent of
$4\pi \sqrt{\vert q^2\vert/2 }$, should be related to the entropy of
these black holes \cite{Russo:1994}. On the other hand the
corresponding black hole solutions were constructed in
\cite{Sen:1994,Sen:1995in} and it was found that their horizon area
vanishes, in contrast with what one would expect on the basis of
the area law.

Of course, higer-order string corrections are expected to modify the
situation at the horizon. One of the ways to incorporate their effect
is to make use of the concept of a `stretched' horizon, a surface
close to the event horizon, whose location has to be chosen carefully
in order that the calculations remain internally consistent. In this
way it is possible to reconcile the non-zero level density with the
vanishing of the classical horizon area \cite{Sen:1995in,Peet:1995},
although the precise proportionality factor in front of $\sqrt{\vert
  q^2\vert/2 }$ cannot be determined.

On the other hand, these corrections will undoubtedly be related to
interactions of higher order in the curvature tensor whose effect can be
studied in the context of the modified entropy formula (\ref{eq:W-entropy})
together with the attractor equations (after all, their derivation did
not require any restriction on the value of the leading contributions
to area and entropy). In fact, some of the results can be read off
easily from the formulae (\ref{eq:nonholostab}) and
(\ref{eq:nonholoentropy}).  They show that the dilaton field becomes
large and real (in contrast with the dyonic case, where the dilaton
could remain finite and complex). Direct substitution yields,
\begin{eqnarray}
\label{eq:entcor-electric}
  S+\bar S &\approx &\sqrt{\vert q^2\vert/2}\,,\nonumber \\ 
  \mathcal{S}_{\text{macro}} &\approx& 
4\,\pi\, \sqrt{\vert q^2\vert/2} -6\, \log{\vert q^2\vert}  \,, 
\end{eqnarray}
where the logarithmic term is due to the non-holomorphic contribution.
Because the dilaton is large in this case, all the exponentials in the
Dedekind eta-function are suppressed and one is at weak string
coupling. Consequently the formalism discussed in
section~\ref{sec:entropy-formula} yields the expected results. In
fact, without the non-holomorphic corrections the result for the
entropy can be obtained directly from (\ref{eq:S-CY-micro}), upon
taking $\hat q_0=q_0$, $C_{ABC}= 0$ and $c_{2\,A}\,p^A = 24\, p^1$.

Not much attention was paid to this particular application of
(\ref{eq:W-entropy}) until recently, when attention focused again
on the electric black holes \cite{Dabholkar:2004yr}, this time
primarily motivated by a reformulation of the black hole entropy
(\ref{eq:W-entropy}) in terms of a mixed partition function
\cite{Ooguri:2004zv}. I turn to the latter topic in the next section.
The observation that (\ref{eq:W-entropy}) can nicely account for the
discrepancies encountered in the classical description of the 1/2-BPS
black holes was first made in
\cite{Dabholkar:2004yr,Dabholkar:2004dq}. Note also that, since the
electric states correspond to perturbative heterotic string states,
their degeneracy is known from string theory and given by
\begin{equation}
  \label{eq:het-string}
  d(q) = \oint {\mathrm d}\sigma\, \frac{{\mathrm
  e}^{\mathrm{i}\pi\sigma q^2}}{\eta^{24}(\sigma)} \approx 
  \exp\left(4\pi\,\sqrt{|q^2|/2} - \tfrac{27}{4}
      \log \vert q^2\vert\right)  \,,
\end{equation}
where the integration contour encircles the point $\exp(2\pi
\mathrm{i} \sigma)=0$. The large-$\vert q^2\vert$ approximation is
based on a standard saddle-point approximation.  Obviously the
leading term of (\ref{eq:het-string}) is in agreement with
(\ref{eq:entcor-electric}). However, the logarithmic corrections carry
different coefficients.
 
At this point I should recall that the dyonic degeneracy formula
(\ref{eq:dvvdeg}) was proposed at the time \cite{Dijkgraaf:1996it} as
an $S$-duality invariant extension of the electric degeneracy formula
(\ref{eq:het-string}). While for the dyonic states the macroscopic
results agree fully with the results obtained from a saddle-point
approximation of (\ref{eq:dvvdeg}), I conclude that the situation
regarding the electric states is apparently more subtle.  Recently,
there have been quite a number of papers about the electric black
holes, discussing the effect of the higher-derivative corrections in
the effective action on the horizon behaviour and on more global
aspects of the black hole solutions
\cite{Sen:2004dp,Hubeny:2004ji,Bak:2005x,Sen:2005x,Dabholkar:2005x};
some of them also discuss the effect of the non-holomorphic
corrections. 

\section{Black hole partition functions}
\label{sec:black-hole-partition}
Recently the study of the BPS black hole entropy received a new
impetous by the conjecture that there exists a black hole partition
function based on a mixed ensemble, which is proportional to the
square of the topological string partition function
\cite{Ooguri:2004zv}. This conjecture was motivated by the observation
that the entropy formula (\ref{eq:W-entropy}) can be rewritten as
\begin{equation}
  \label{eq:real-F}
  \mathcal{S}_{\text{macro}}(p,q) = \mathcal{F}(\phi,p) - \phi^I\, 
  q_I \,,\quad \text{with} \quad 
  q_I=   \frac{\partial\mathcal{F}(\phi,p)}{\partial \phi^I} \,.
\end{equation}
Here the $Y^I$ are expressed in terms of the magnetic charges $p^I$
and (real) electrostatic potentials $\phi^I$ at the horizon,
\begin{equation}
\label{eq:electro-phi} 
  Y^I = \frac{\phi^I}{2\pi} + \frac{\mathrm{i} p^I}{2} \;,
\end{equation}
and the real function $\mathcal{F}(\phi,p)$ is defined by 
\begin{equation}
  \label{eq:def-real-F}
  \mathcal{F}(\phi,p) = 4\pi \,\mathrm{Im}
  [\,F(Y,\Upsilon)]_{\Upsilon=-64} \,, 
\end{equation}
so that the magnetic attractor equations are already imposed and the
electric ones correspond to the second equation in (\ref{eq:real-F}).
The relation (\ref{eq:real-F}) constitutes precisely a Legendre 
transform and this fact motivated the introduction of a mixed black
hole partition function of the form,
\begin{equation}
  \label{eq:mixed-partition}
\mathrm{e}^{\mathcal{F}(\phi,p)} = 
Z_{\text{BH}}(\phi,p)  = \sum_{\{q_I\}} \; d(q,p)\, 
\mathrm{e}^{ q_I \,\phi^I} \,,  
\end{equation}
where $d(q,p)$ denotes the microscopic black hole degeneracies. This
partition function is called a `mixed' partition function, as it treats
the electric and the magnetic charges differently: with respect to the
magnetic charges one is dealing with a microcanonical ensemble, and
with respect to the electric charges one has a canonical ensemble.
Note that the left-hand side of (\ref{eq:mixed-partition}) can be
written as the modulus square of $\exp[-2\pi\mathrm{i}
F(Y,\Upsilon)]$, where the $Y^I$ are given by (\ref{eq:electro-phi})
and $\Upsilon=-64$. The holomorphic expression $\exp[-2\pi\mathrm{i}
F(Y,\Upsilon)]$ is actually related to the partition function for the
topological string; the non-holomorphic corrections, which we
suppressed here, are related to the so-called holomorphic anomaly
\cite{BCOV}.

The equation (\ref{eq:mixed-partition}) implies that the black hole
degeneracies can be expressed as a Laplace transform of the partition
function $Z_{\text{BH}}(\phi,p)$,
\begin{equation}
  \label{eq:laplace}
  d(q,p) \sim \int \prod_I \mathrm{d}\phi^I \, 
  \left\vert\mathrm{e}^{-2\pi\mathrm{i} F(Y,\Upsilon)}\right \vert^2 \; 
   \mathrm{e}^{ - q_I \,\phi^I} \,.  
\end{equation}
where the $Y^I$ are still given by (\ref{eq:electro-phi}) and
$\Upsilon=-64$.  For large values of the $q_I$ the Laplace transform
can be solved by a saddle-point approximation which leads to the
exponent of the entropy $\mathcal{S}_{\text{macro}}(p,q)$ in accord
with (\ref{eq:real-F}). It turns out that there exists a variety of
these integral representations. They belong to a certain
hierarchy and are generically related through a series of
saddle-point approximations \cite{CdWKM2}.

These integrals are, however, not properly defined. Leaving the
question of convergence aside, a crucial issue concerns
electric/magnetic duality. The attractor equations
(\ref{eq:attractor}) and the expression for the macroscopic entropy
(\ref{eq:W-entropy}) are manifestly consistent with this duality.
Therefore the expression for the entropy transforms as a scalar
function and its expression in a dual description is simply obtained
by applying the duality on the charges $p^I$ and $q_I$, {\it i.e.},
$\mathcal{S}^\prime_{\text{macro}}(p^\prime,
q^\prime)=\mathcal{S}_{\text{macro}}(p,q)$. This implies, in
particular, that the entropy will be invariant under any subgroup of
the electric/magnetic duality group that constitutes an invariance of
the model in question. The same property holds presumably for
the microscopic black hole degeneracies, $d(q,p)$.  Indeed, this is 
nicely demonstrated for the heterotic black holes, where both the
macroscopic and the microscopic description were invariant under $T$-
and $S$-duality. On the other hand, the mixed partition function and
the functions $F(Y,\Upsilon)$ and $\mathcal{F}(\phi,p)$ do not
transform as functions under electric/magnetic duality; this is
already obvious from the fact that the $(p^I,\phi^I)$ do not transform
simply under this duality, unlike $(q_I,p^I)$. Hence it is not
surprising that a straighforward integration leads to results that do
not respect the various symmetries (as was, for instance, noted in
\cite{Dabholkar:2005x}). Clearly the integration requires an
appropriate measure in order to yield invariant results. In the
context of the aforementioned hierarchy, it turns out that it is
possible to construct such a measure \cite{CdWKM2}.

So far the integral representation based on (\ref{eq:laplace}) has
mainly been studied for the 
(electric) 1/2-BPS states (sometimes called `small' black holes, as
their horizon area vanishes at the classical level)
\cite{Dabholkar:2004yr,Sen:2005x,Dabholkar:2005x}. It seems
reasonable to expect that there exists a single integral formula that
pertains to both electric and dyonic black holes. However, it turns
out that there are crucial differences when evaluating the
corresponding integrals. To illustrate some of these subtleties, let
me postulate the following integral expression for heterotic black
holes,
\begin{equation}
  \label{eq:dilatonic-integral}
  d(p,q) = \int \frac{{\mathrm{d}^2S}}{(S+\bar S)^2} \;
  \mathrm{e}^{\mathcal{S}_{\mathrm{macro}}}\;,
\end{equation}
where $\mathcal{S}_{\mathrm{macro}}$ is given by
(\ref{eq:nonholoentropy}). Because of the factor $(S+\bar S)^{-2}$ the
integral is invariant under $S$-duality and it is also invariant under
$T$-duality. While the above expression is not completely unfounded, as we
shall see, it cannot be entirely correct. For simplicity let me suppress
the exponential terms in the Dedekind eta-function, so that
\begin{equation}
  \label{eq:S-noninstanton}
  \mathcal{S}_\text{macro} = 
   - \pi \left[ \frac{q^2 - \mathrm{i} p\!\cdot\! q \, 
  (S - {\bar S}) + p^2 \,|S|^2} {S + {\bar S}} \right] 
  + 2\pi (S+\bar S)  -12 \, \log(S + {\bar S})  \;,
\end{equation}
and substitute them into the integral
(\ref{eq:dilatonic-integral}). Note that in this approximation we
have lost the manifest $S$-duality invariance.

For the electric case where $p^2$ and $p\cdot q$ vanish, the integrand
depends only on the real part of the dilaton field. Because of the
$S$-duality invariance of the exact integral
(\ref{eq:dilatonic-integral}), the integral over the imaginary part
extends only over a finite interval (in other words, we can restrict
the integration to a suitable fundamental domain), so the only
nontrivial integral is the one over the real part of $S$. Denoting the
latter by $\tfrac{1}{2}x$, the integral takes the form (up to a
numerical factor),
\begin{equation}
  \label{eq:electric-dilatonic-integral}
  d(p,q) \propto \int \frac{{\mathrm{d}x}}{x^{14}} \;
  \exp\Big[ - \frac{\pi\,q^2}{x} + 2\pi\,x\Big] \;,
\end{equation}
which equals a modified Bessel function of a certain degree (which was
already noted in \cite{Dabholkar:2005x}), whose asymptotic behaviour for
large negative $q^2$ coincides precisely with the microscopic result
(\ref{eq:het-string}). The asymptotic result follows straightforwardly
from a saddle-point approximation of
(\ref{eq:electric-dilatonic-integral}), which shows that the dilaton
field at the saddle point is large and given by the first equation of
(\ref{eq:entcor-electric}). The semi-classical correction which is
included in the saddle-point approximation, thus changes the
logarithmic term $-6 \log \vert q^2\vert$ in the second
equation of (\ref{eq:entcor-electric}) into the term $-\tfrac{27}4
\log \vert q^2\vert$ in (\ref{eq:het-string}). This is a gratifying result.

One can repeat the same exercise for the dyonic case. Then the
integrand does depend on both the real and imaginary parts of $S$ and
the integral over the imaginary part now takes the form of a Gaussian
integral, which can be performed explicitly. Subsequently one is again
left with an integral over $x$ (after a suitable rescaling of the
integration variables) which reads as follows (suppressing again a
numerical factor),
\begin{equation}
  \label{eq:dyon-dilatonic-integral}
  d(p,q) \propto \int \frac{\mathrm{d}x}{x^{27/2}} \;
  \exp\Big[ - \frac{\pi\,[p^2q^2-(p\cdot q)^2]}{x} - \pi\Big(\frac14- 
  \frac2{p^2}\Big) x\Big] \;.
\end{equation}
This result also takes the form of a modified Bessel function, but now
of a different degree. The asymptotic behaviour follows from a
saddle-point approximation, which leads to the entropy
(\ref{eq:class-het-entropy}) modified by a logarithmic term.  The
coefficient of the logarithm is, however, not in agreement with
(\ref{eq:nonholostab}) and (\ref{eq:nonholoentropy}), which, as I
discussed in section~\ref{sec:heter-black-holes}, have been shown to
follow from the formula (\ref{eq:dvvdeg}) for the dyonic degeneracies,
up to inverse powers of the charges \cite{CDWKM}. Although
(\ref{eq:dilatonic-integral}) thus seems to lead to a desirable result
in the electric case, it fails to do so in the dyonic case.  This does
not yet pose a problem as (\ref{eq:dilatonic-integral}) was only
introduced as an example to illustrate typical features of these
integral representations. In certain cases this particular formula
does play a role at an intermediate stage of a more complete
derivation, but the latter reveals that it is actually the electric
case where most of the subtleties arise, while the dyonic case is much
more generic, at least when one adopts an appropriate measure for the
integral \cite{CdWKM2}.

In spite of these open questions it is clear that surprising progress
has been made in the understanding of black hole entropy. The
intriguing connection with topological string theory and the interplay
with supergravity and general relativity open up new perspectives
for research. I expect that this will remain an exciting subject for
study in the years to come.

\vspace{4mm}
\noindent
Most of my own work on the topic of this talk has been in
collaboration with Gabriel Lopes Cardoso, J\"urg K\"appeli and Thomas
Mohaupt whom I thank for valuable comments on the text. It is a
pleasure to acknowledge the kind hospitality of the Yukawa Institute
(Kyoto University) where most of this contribution was prepared. This
work is partly supported by EU contract MRTN-CT-2004-005104.

\providecommand{\href}[2]{#2}
\begingroup\raggedright \endgroup

\end{document}